 \documentclass[review,12pt]{elsarticle}

\usepackage{amssymb,amsmath}
\usepackage{array}
\usepackage{multirow}
\usepackage{multicol}
\usepackage{tabularx}
\usepackage{colortbl}
\usepackage{nicefrac}
\usepackage{mathrsfs}
\usepackage{verbatim}
\usepackage[T1]{fontenc}

\usepackage[squaren, binary]{SIunits}

\newcommand*\diff{\mathop{}\!\mathrm{d}}

\newcolumntype{Z}{>{\centering\arraybackslash}X}

\biboptions{compress}

\DeclareTextAccent{\myacc}{T1}{4}

\usepackage{lineno}

\biboptions{compress}

\journal{Nuclear Instruments and Methods in Physics Research Section A}

\begin{document}

\begin{frontmatter}

\title{Characterization and performances of a monitoring ionization chamber dedicated to IBA-universal irradiation head for Pencil Beam Scanning.}

\author[1]{C. Courtois}
\author[1]{G. Boissonnat\corref{cor1}}
\ead{boissonnat@lpccaen.in2p3.fr}
\author[2]{C. Brusasco}
\author[1]{J. Colin}
\author[1]{D. Cussol}
\author[1]{JM. Fontbonne}
\author[2]{B. Marchand}
\author[2]{T. Mertens}
\author[2]{S. de Neuter}
\author[1]{J. Peronnel}

\address[1]{LPC (IN2P3-ENSICAEN-UNICAEN), 6 Boulevard Mar\'echal Juin, 14050 Caen, France}
\address[2]{IBA, 3 Chemin du Cyclotron, 31348 Louvain-la-Neuve, Belgium}
\cortext[cor1]{Corresponding author. Tel.: +33 2 31 45 25 30 }

\begin{abstract}
 Every radiotherapy center has to be equipped with real-time beam monitoring devices. In 2008, we developed an Ionization Chamber in collaboration with the IBA company (Ion Beam Applications). This monitoring device called IC2/3 was developed to be used in IBA universal irradiation head for Pencil Beam Scanning (PBS).
Here we present the characterization of the IC2/3 monitor in the energy and flux ranges used in protontherapy. The equipment has been tested with an IBA cyclotron able to deliver proton beams from 70 to 230~MeV.
This beam monitoring device has been validated and is now installed at the Westdeutsches Protonentherapiezentrum Essen protontherapy center (WPE, Germany). The results obtained in both terms of spatial resolution and dose measurements are at least equal to the initial specifications needed for PBS purposes. The detector measures the dose with a relative uncertainty lower than $1\%$ in the range from 0.5~Gy/min to 8~Gy/min while the spatial resolution is better than 250~$\mu$m. The technology has been patented and five IC2/3 chambers were delivered to IBA. Nowadays, IBA produces the IC2/3 beam monitoring device as part of its Proteus 235 product.

\end{abstract}
\end{frontmatter}
 
\section{Introduction}
\label{sec_intro}

Protontherapy \cite{ICRU,Chu,Scardt} is a special kind of radiotherapy using proton beams with an energy ranging between 70~MeV and 230~MeV. Because proton beams undergo little scattering in matter and deliver most of the irradiation dose at a specific depth (Bragg Peak) depending on their initial energy, tumors can be precisely targeted while sparing surrounding tissues. At first, protontherapy was mainly limited to very specific kind of cancers, such as eye cancers, but with time its applications got more and more diversified to become fully part of the cancer therapy toolbox.
While the energy dose delivered by the beam to the tumor causes its inactivation, the dose prescribed by the physician and the one delivered by the irradiation unit should be as close as possible. One of the main issues is to conform the delivered dose to the target volume with no consideration on its shape. To this end, the IBA company and the Massachusetts General Hospital team have developed a dose delivery mode called Pencil Beam Scanning (PBS) \cite{Marchand}. It uses a very narrow beam (few millimeters of standard deviation) to scan (similarly to spot scanning \cite{Lomax}) the whole tumor in three dimensions (x and y by scanning and z by changing the beam energy) while modulating the beam intensity to get the prescribed dose map. One of the fundamental safety requirements of the PBS technology is to check in real-time during the scanning process that the beam is at the right position in space with the correct intensity.
In this work we are going to focus on the charge measured by an air-ionization chamber (or IC) $Q_{IC}$ (in pC) while the final aim of this detector is to have an estimate of the deposited the Dose $D$ (in Gy). Those quantities are deeply linked  (see Eq.~\ref{EqDose},~\ref{EqCharge},~\ref{EqDose2}) through the proton fluency $\Phi$ (in mm$^{-2}$), the linear energy transfer $LET$ and $LET_{IC}$ (MeV\cdot mm$^{-1}$) respectively in the medium where the dose is deposited and in the detector, the ionization potential in air ($W_{air}$ in MeV), the air-gap $d$ (in mm) of the IC and also the density of the medium in which the dose is deposited $\rho$ (in g\cdot cm$^{-3}$).

\begin{equation}
D=1.6\times 10^{-7}\times \Phi \times \frac{LET}{\rho}
\label{EqDose}
\end{equation}

\begin{equation}
Q_{IC}=1.6\times 10^{-10}\times \Phi \times d\times \frac{LET_{IC}}{W_{air}}
\label{EqCharge}
\end{equation}

\begin{equation}
D=10^{3}\times Q_{IC} \times \frac{LET}{LET_{IC}} \times  \frac{W_{air}}{\rho \times d}
\label{EqDose2}
\end{equation}

In practice, every ionization chamber has to be carefully calibrated, to allow the conversion from $Q_{IC}$ to $D$, through a commissioning process. 

\section{Material and methods}
\subsection{Description of the IC2/3 beam monitor}
\begin{figure}[!ht]
\begin{center}
\includegraphics[angle=0, width=0.8\columnwidth]{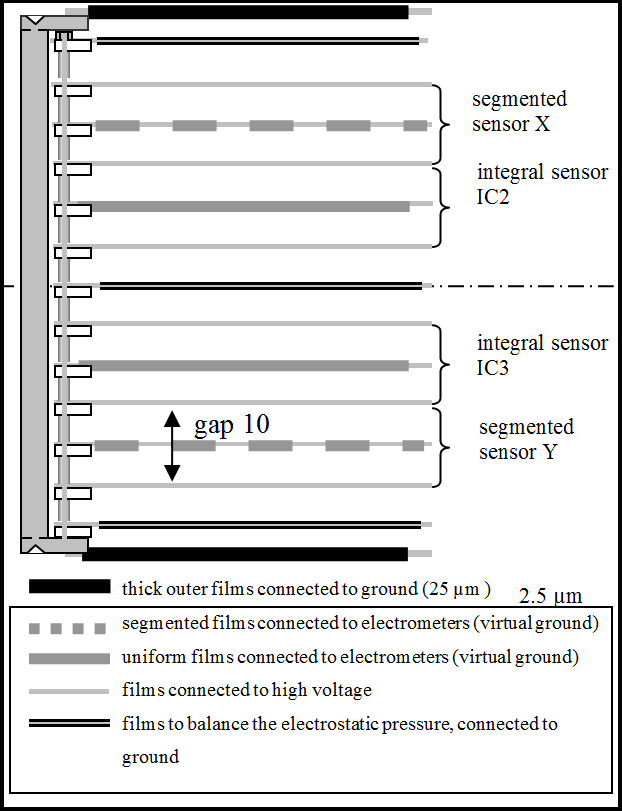}
\end{center}
\caption{Vertical section of IC2/3.}
\label{ICmapping}
\end{figure}

We have developed a 320~mm $\times$ 320~mm parallel plate ionization chamber composed of fifteen Mylar foils separated by 5~mm air gaps. The detector is called IC2/3 for Ionization Chamber 2 and 3 while it is composed of two identical and fully independent (power supply and electronic acquisition setup) units IC2 and IC3. These two chambers allow achieving dose measurement redundancy, a fundamental requirement for safety purposes. As sketched in Fig.~\ref{ICmapping} each unit is composed of five 2.5~$\mu$m Mylar electrodes coated on both sides with aluminum or gold. Three are connected to the high voltage (or HV) while the two others are measurements electrodes (virtually grounded), one being used for dose measurement (uniform film) and the second one for beam position measurement (striped film) along one axis (horizontal for IC2 and vertical for IC3). Apart from the two units, three other films are put to the ground to ensure the electrostatic pressure equilibrium. In addition, two thicker (25~$\mu$m) Mylar film are used to cover both entrance and exit windows. The whole chamber is 6.86~cm thick for a total water equivalent thickness of 187~$\mu$m.

The specifications for the IC2/3 monitoring device were to have a global uniformity in the measurement dose response better than 1\% after correction (inhomogeneities, edge effects,...) .

Because the beam spot swipes the whole detector, the detector response should be position-independent. The beam-induced ionization in the air-detector should be identical at any position in the detector; while the beam direction has a normal incidence with the detector, the main parameter affecting the response is the potential variation of the 5mm air gap between electrodes (see Eq.~\ref{EqCharge}). The design study showed that the layers localized at the center of the detector are less affected by the electrostatic pressure and therefore have less variation in their gap width. As a consequence, the dose measurement layers are located in the middle of the chamber and are surrounded by the position measurement Mylar films. 

To localized the beam centroid, there are two main solutions, films can be either pixalized or striped \cite{Giordanengo} for the position measurement layer. Here, it was chosen to use strips to have less electronic channels at identical spatial resolution. The position measurement layers are plain Mylar films covered by Au-strips on both sides deposited by vacuum evaporation. There are 64 4.8~mm-stripes with a 5~mm-step, oriented to give the x and y position for IC2 and IC3 respectively. 

Dose measurement layers are single layer aluminized Mylar foils covered in the plain side by a thin gold layer (200~nm). The dose is measured by integrating the signal collected on each film at a frequency of 2~kHz (allowing a time resolution of 500~$\mu$s) with the Faster acquisition framework (developed at LPC-Caen) using the electrometer DDC232 from $Texas~Instrument$).

\subsection{Set up}

Three series of tests were performed at the Westdeutsches Protontherapiezentrum Essen to evaluate the performances of the IC2/3 detector with a proton beam. The WPE is equipped with an IBA Proteus 235 device, a cyclotron producing proton beams in the energy range from 70 to 230~MeV. The tests were realized in a treatment room with a Pencil Beam Scanning delivery technique and equipped with an isocentric gantry.
Results of the first two IC2/3 prototypes were compared to those of a PTW ionization chamber ($Physikalisch-Technische Werkst$\myacc{a}$tten$ - $Roos Electron Ionization Chamber$) placed in air one meter away from the isocenter. During the tests different beam configurations (Table~\ref{table:Beams}) were used with different energies, intensity and sizes, while the beam used is  a Gaussian shape its size (standard deviation) will be referred to as $\sigma _{beam}$.

\begin{table}[!ht]
\caption{Beam configurations.}
\centering
\footnotesize
\label{table:Beams}
\renewcommand{\arraystretch}{1.3}
{
\begin{tabularx}{\linewidth}{ZZZZ}
\hline
\hline
 & Energy & Beam Intensity & $\sigma _{beam}$\\
\hline
1&100~MeV& 1~nA & 4.76~mm\\
2&100~MeV& 2.8~nA& 4.76~mm\\
3&230~MeV& 2~nA& 2.85~mm \\
4&230~MeV&20~nA& 2.85~mm\\
5&150~MeV& 0.5-3~nA& 3.50~mm\\
\hline
\hline
\end{tabularx}}
\end{table}

\subsection {Repeatability Measurement}
To ensure that the detector is working properly, the first step, it to do multiple measurements in the same condition. Here, the aim is to get a standard deviation on the repeatability measurements (on uniform films) better than 1\%. Measurements were performed at an intensity of 2.8~nA with 100~MeV-protons and a $\sigma _{beam}$= 4.76~mm while the HV of IC2/3 was set at 1.2~kV. The charge measured by both IC2 and IC3 (respectively $Q_{IC2} $ and $Q_{IC3}$) were realized at the same time and normalized by the PTW measurement ($Q_{ref} $) to prevent from beam fluctuations.

\subsection {Charge collection efficiency}
Positive and negative charges created in the chamber by protons might recombine while traveling toward their respective collection electrodes \cite{Boag}. This effect depends, among other factors, on the gas density and the drifting time of each charge type. The saturation of the operating mode is reached when all charges created are collected (100\% collection efficiency) and this can be approached when the polarization voltage of the electrodes is high enough. The goal was to determine the lowest polarization voltage allowing a collection efficiency better than 99\%. Measurement were performed irradiating the detector with a beam at constant current and decreasing the polarized voltage by steps of 100~V. Decreasing the voltage instead of increasing it allows preventing recombination to occur at some measurement and modifying results of the following steps. This way, only the last measurement (at lowest voltage) might be perturbed. This saturation curves were measured for the four first beam configurations of Table~\ref{table:Beams}.

\subsection {Linearity}

The aim of the linearity test is to check that the relation between the deposited dose and the measured charge is truly linear. Measurements were performed by irradiating the detector with a decreasing beam intensity from 3~nA to 0.5~nA (beam configuration 5 on the Table~\ref{table:Beams}) while keeping the integration time constant. In order to reduce the beam intensity in the PTW reference chamber and prevent it from any saturation, it was placed (in air) one meter away from the IC2/3 detector with a 3~mm thick lead foil between them enlarging the beam by scattering. This test would be consider satisfactory if the standard deviation on the three ratios:  $Q_{IC2}/Q_{Ref} $, $Q_{IC3}/Q_{Ref} $ and $Q_{IC2}/Q_{IC3} $ is lower than 1\%.

\subsection {Uniformity of the detector response}
\begin{figure}[!ht]
\begin{center}
\includegraphics[angle=0, width=\columnwidth]{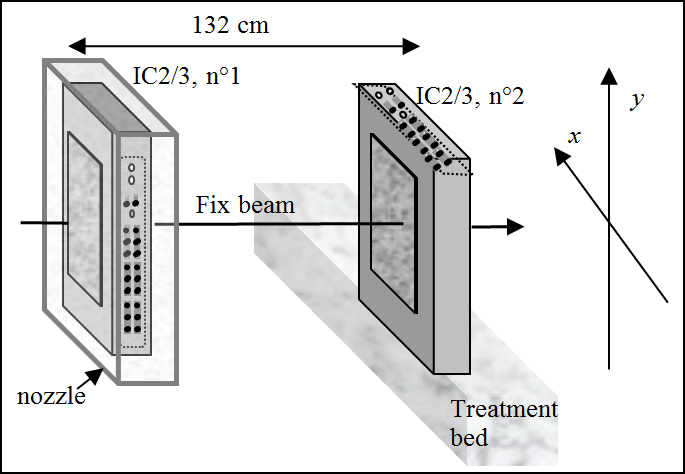}
\end{center}
\caption{Set up for the uniformity test.}
\label{UniformitySchem}
\end{figure}

The detector can measure either the beam intensity integrated on the whole detector area or on each strip, allowing to infer beam position and size. The uniformity test allows testing that the intensity is correctly measured on the whole detector area. As the required dose uniformity should be less than 1\%, the mechanical structure should insure that the variation in the air gap (see Eq.~\ref{EqCharge}) has to be under 1\%. The experimental setup to measure such uniformity consists of two IC2/3 chambers (Fig.~\ref{UniformitySchem}). The first one is fixed on the nozzle and the second one is located 132~cm away on a mobile treatment bed. Using two IC2/3 chambers allows to avoid the beam inhomogeneities by dividing the charge measured at a given position in the chamber 2 ($Q_2(x,y)$) by the one measured at the center of chamber 1 ($Q_1(0,0)$). This test was performed with the beam configuration 3 on the Tab.~\ref{table:Beams}.
 
\subsection {Spatial Resolution}

Each strip of the chamber measures a part of the charge created in the air volume crossed by the beam. In the orthogonal plane to the beam, the beam distribution exhibits a two dimensional Gaussian shape. Therefore the signal integrated by each strip corresponds to this Gaussian curve integrated over the strip surface. Matching an unknown Gaussian curve (Eq.~\ref{EqGauss}) with the charge distribution measured by the two arrays of strips allows finding the height ($h_{fit}$), the center ($\mu_x$, $\mu_y$) and the standard deviation ($\sigma_{beam}$) of the Gaussian in two dimensions for the fit. The Gaussian fit is still possible on the edge of the chamber even if the measured dose profile is not perfectly Gaussian as shown in the previous paragraph. 

\begin{equation}
fit(x_i)=h_{fit}\int_{x_i-0.25}^{x_i+0.25}\frac{1}{\sqrt{2\pi}\sigma_{beam}}exp \left( -\frac{1}{2}\left(\frac{\mu_x-x}{\sigma_{beam}}\right )^2\right )\diff x \\
\label{EqGauss}
\end{equation}

Even if the curve fitting is mandatory to get values of $\sigma_{beam}$ and $h_{fit}$, in practice, a faster way to get a good estimate ($\tilde{\mu}_x$) of the centroid position ($\mu_x$) is to use barycentric calculation, as presented on Eq. \ref{EqBarycentre} where $x_i$ and  $Q_i$ are respectively the central position on the  $i$-strip and the charge collected on it. We can calculate of an estimated error ($|\tilde{\mu}_x-\mu_x|$) as a function of the relative position of the centroid on its strip ($|x_i-\tilde{\mu}_x|$) for different beam sizes, the results are presented in term of resolution in Table~\ref{table:CentroidEstimate}. Therefore, to get a resolution better than 5~$\mu$m the beam standard deviation should be larger than 2.5~mm, and this is the case with every beam used in this article.

\begin{equation}
\tilde{\mu}_x=\frac{\sum{ Q_i x_i }}{\sum{ Q_i}}
\label{EqBarycentre}
\end{equation}


\begin{table}[!ht]
\caption{Resolution obtained for barycentric calculations as a function of the beam size.}
\centering
\footnotesize
\label{table:CentroidEstimate}
\renewcommand{\arraystretch}{1.3}
{
\begin{tabularx}{\linewidth}{ZZZZZ}
\hline
\hline
$\sigma_{beam}$ & 1.5~mm &2~mm &2.5~mm &3~mm\\
\hline
Resolution & 120~$\mu$m &30.2~$\mu$m &5.12~$\mu$m &0.584~$\mu$m \\
\hline
\hline
\end{tabularx}}
\end{table}

To measure the resolution (using beam configuration 5 in Table~\ref{table:Beams}) the variation of beam centroid calculation is measured over time with a fixed beam and its standard deviation should be better than 250~$\mu$m (as required by IBA).
After that, the impact of the relative centroid estimate $|\tilde{\mu}_x-x_i|$) on the $\sigma_{beam}$ calculation will be measured with the same beam but in scanning mode to evaluate the resolution on the beam size.

\section{Results}

\subsection {Repeatability Measurement}
The repeatability measurements $Q_{IC2}/Q_{Ref} $, $Q_{IC3}/Q_{Ref} $ and $Q_{IC2}/Q_{IC3}$ are presented normalized in Table~\ref{table:Repeatability}. The measured relative standard deviation is better than 0.6\% in good agreement with the desired specifications of 1\%. 

\begin{table}[!ht]
\caption{Repeatability test measurements made on the uniform films of IC2/3 corrected from beam fluctuation by the PTW reference chamber and normalized by their mean values (2.8~nA, 100~MeV, $\sigma _{beam}$= 4.76~mm and 1.2~kV).}
\centering
\footnotesize
\label{table:Repeatability}
\renewcommand{\arraystretch}{1.3}
{
\begin{tabularx}{\linewidth}{ZZZZ}
\hline
\hline
Series & $\frac{Q_{IC2}}{Q_{ref}}$(\%) & $\frac{Q_{IC3}}{Q_{ref}} $(\%) & $\frac{Q_{IC2}}{Q_{IC3}}$(\%)\\
\hline
1	&100.069	&100.081	&99.989\\
2	&99.842	&99.850	&99.993\\
3	&100.676	&100.655	&100.022\\
4	&100.073	&100.073	&100.001\\
5	&100.748	&100.731	&100.018\\
6	&100.679	&100.664	&100.016\\
7	&99.096	&99.072	&100.025\\
8	&100.248	&100.221	&100.028\\
9	&100.077	&100.083	&99.995\\
10	&99.214	&99.207	&100.008\\
11	&99.292	&99.384	&99.908 \\
\hline
Relative standard deviation (\%)	&0.593	&0.580&	0.033\\
\hline
\hline
\end{tabularx}}
\end{table}

\subsection {Charge collection efficiency}

\begin{figure}[!ht]
\begin{center}
\includegraphics[angle=0, width=\columnwidth]{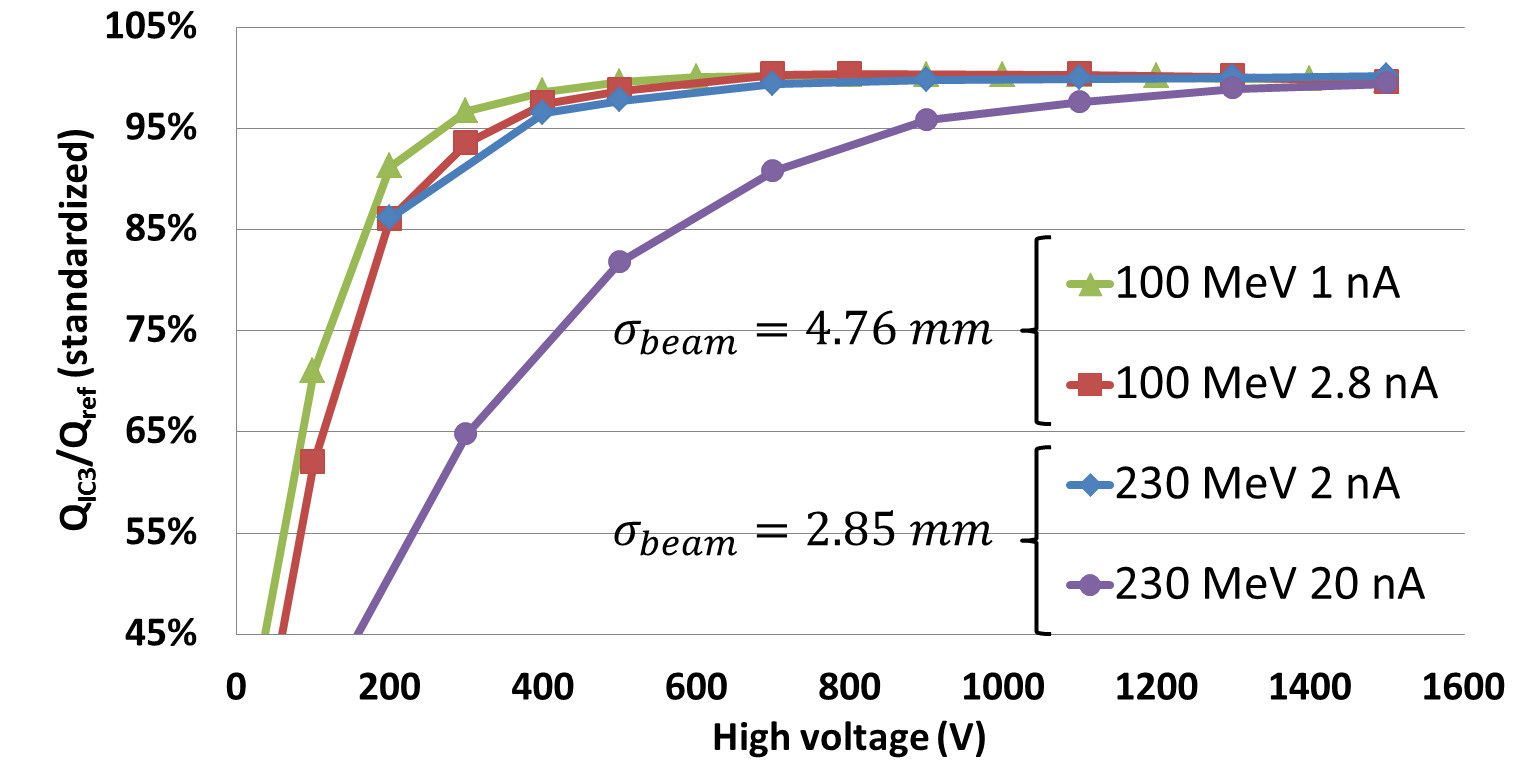}
\includegraphics[angle=0, width=\columnwidth]{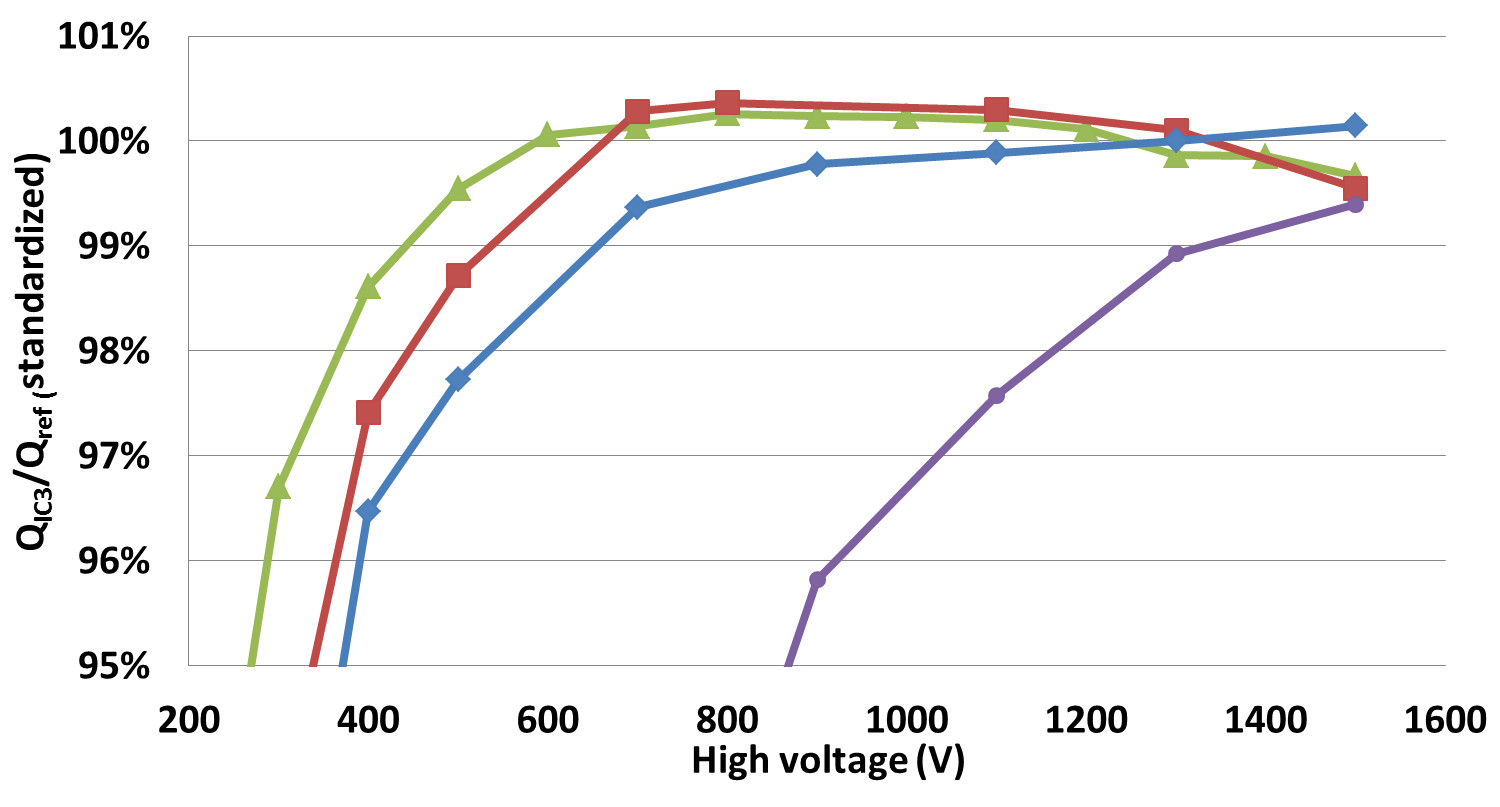}
\end{center}
\caption{ $Q_{IC}//Q_{Ref} $ normalized for four beams configurations.}
\label{SatCurve}
\end{figure}

Measurements (with the four beam configurations) were done on both uniform films from IC2 and IC3, the results $Q_{IC3}/Q_{Ref} $ as a function of the voltage for IC3 are presented (Fig.~\ref{SatCurve}), normalized by the mean value of the measurement above 1.2~kV (excepted for the 230~MeV - 20~nA measurement). A $Q_{IC3}/Q_{Ref}$ ratio below 100\% means that not every charge created is collected and that recombination still occurs. For a high beam intensity (20~nA), a 1.5~kV potential difference ensure a collection efficiency above 99.5\%. With a lower beam current, the ionization density decreases and so the probability of two charges of opposite sign to recombine is even lower and the polarization voltage needed to reach the saturation point could be lower than 1.5~kV. With a standard beam condition (100~MeV, 2.8~nA), 1.2~kV is enough to get 99.5\% collection efficiency.

A $Q_{IC3}/Q_{Ref}$ ratio below 100\% means that not every charge created is collected and that recombination still occurs. For a high beam intensity (20~nA), a 1.5~kV potential difference ensure a collection efficiency above 99.5\%. With a lower beam current, the ionization density decreases and so the probability of two charges of opposite sign to recombine is even lower and the polarization voltage needed to reach the saturation point could be lower than 1.5~kV. With a standard beam condition (100~MeV, 2.8~nA), 1.2~kV is enough to get 99.5\% collection efficiency.

\begin{figure}[!ht]
\begin{center}
\includegraphics[angle=0, width=\columnwidth]{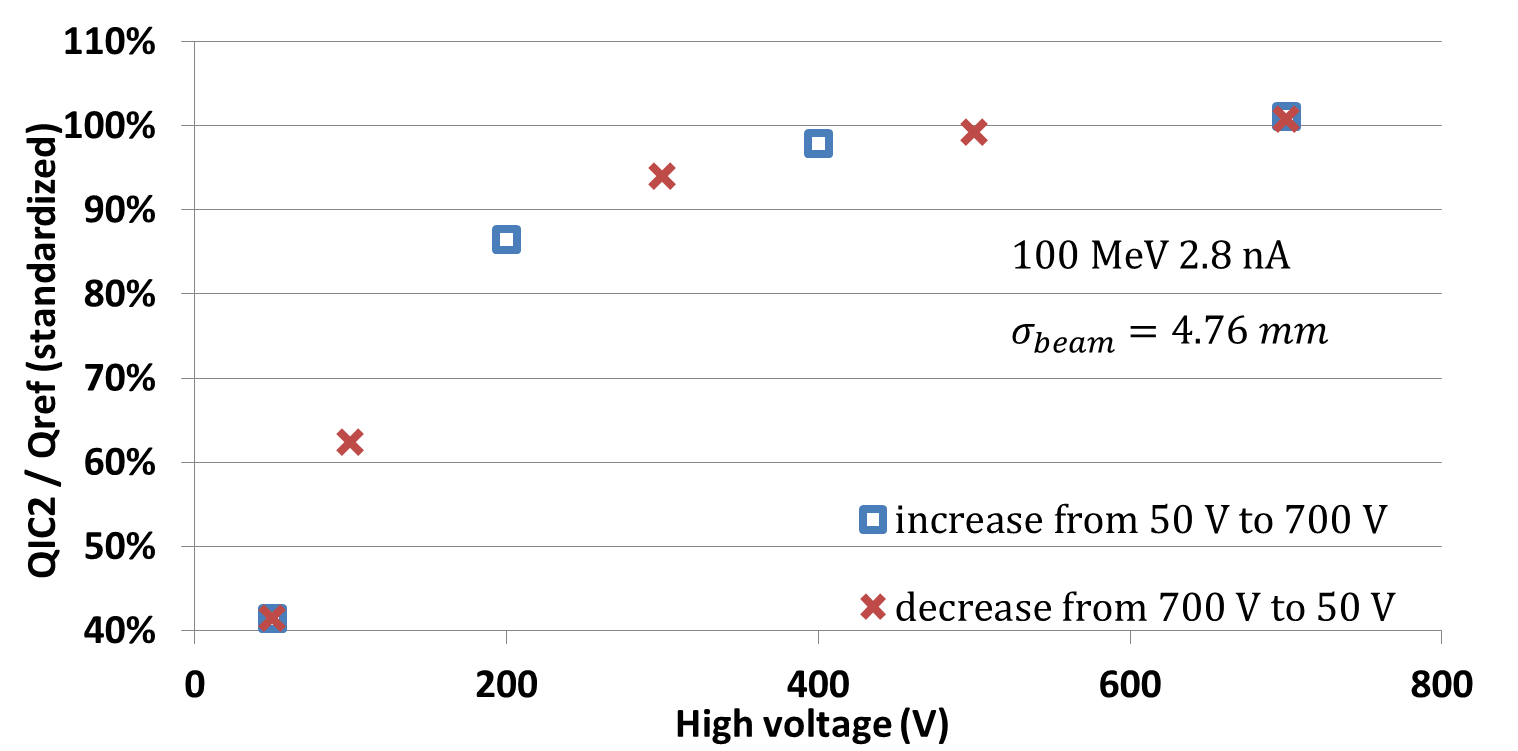}
\end{center}
\caption{Test to evaluate the hysteresis phenomenon at 2.8~nA and 100~MeV.}
\label{hysteresis}
\end{figure}

Fig.~\ref{hysteresis} presents the $Q_{IC2}/Q_{Ref} $ ratio measured while doing an hysteresis test. No noticeable difference can be seen between measurements at same HV.

\subsection {Linearity}

\begin{figure}[!ht]
\begin{center}
\includegraphics[angle=0, width=\columnwidth]{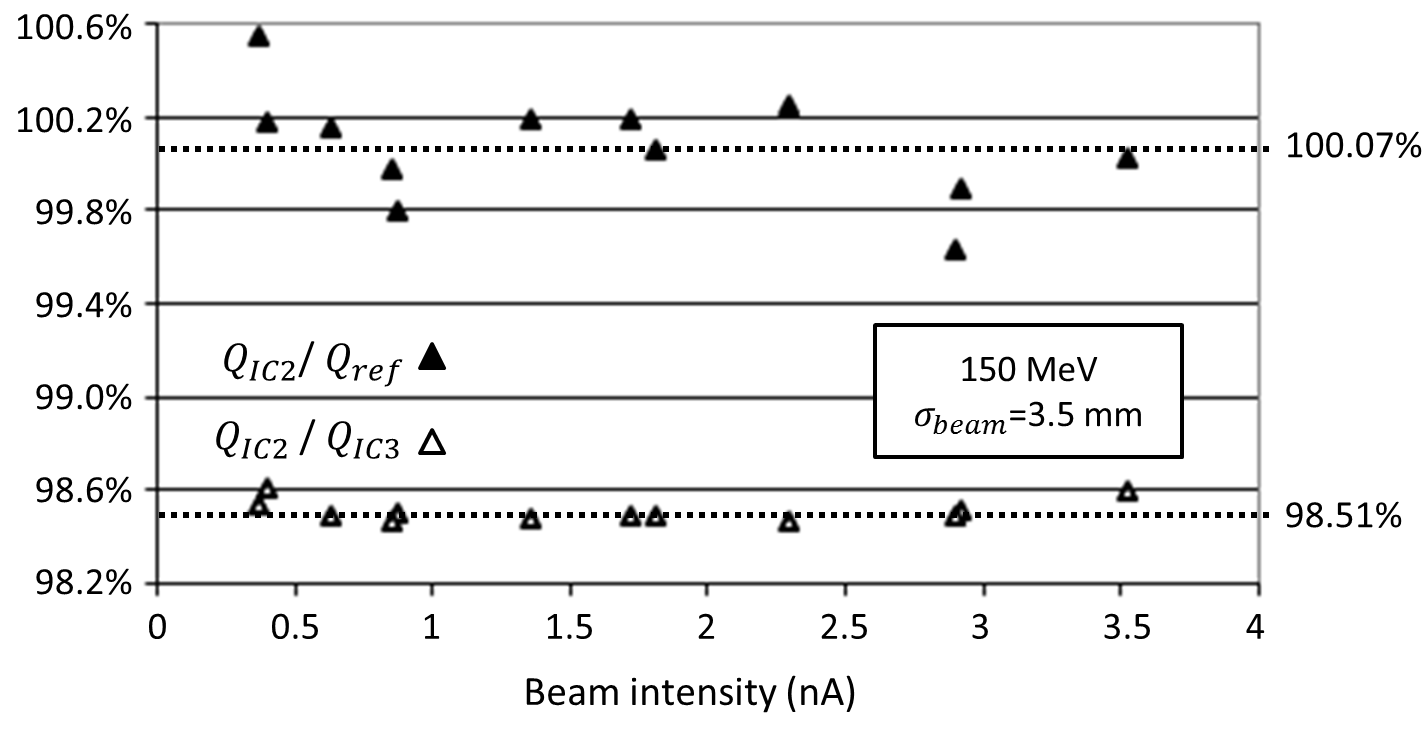}
\end{center}
\caption{ $Q_{IC2}/Q_{Ref} $ and $Q_{IC2}/Q_{IC3} $ versus the beam intensity.}
\label{Linearity}
\end{figure} 

Fig.~\ref{Linearity}, the ratio $Q_{IC2}/Q_{IC3}$ is nearly constant with a mean value of 98.51\% and a standard deviation of 0.047\%. The curve $Q_{IC2}/Q_{Ref}$ shows more fluctuations with a standard deviation of 0.24\% but still under the 1\% required.

\subsection {Uniformity of the detector response}

 \begin{figure}[!ht]
\begin{center}
\includegraphics[angle=0, width=\columnwidth]{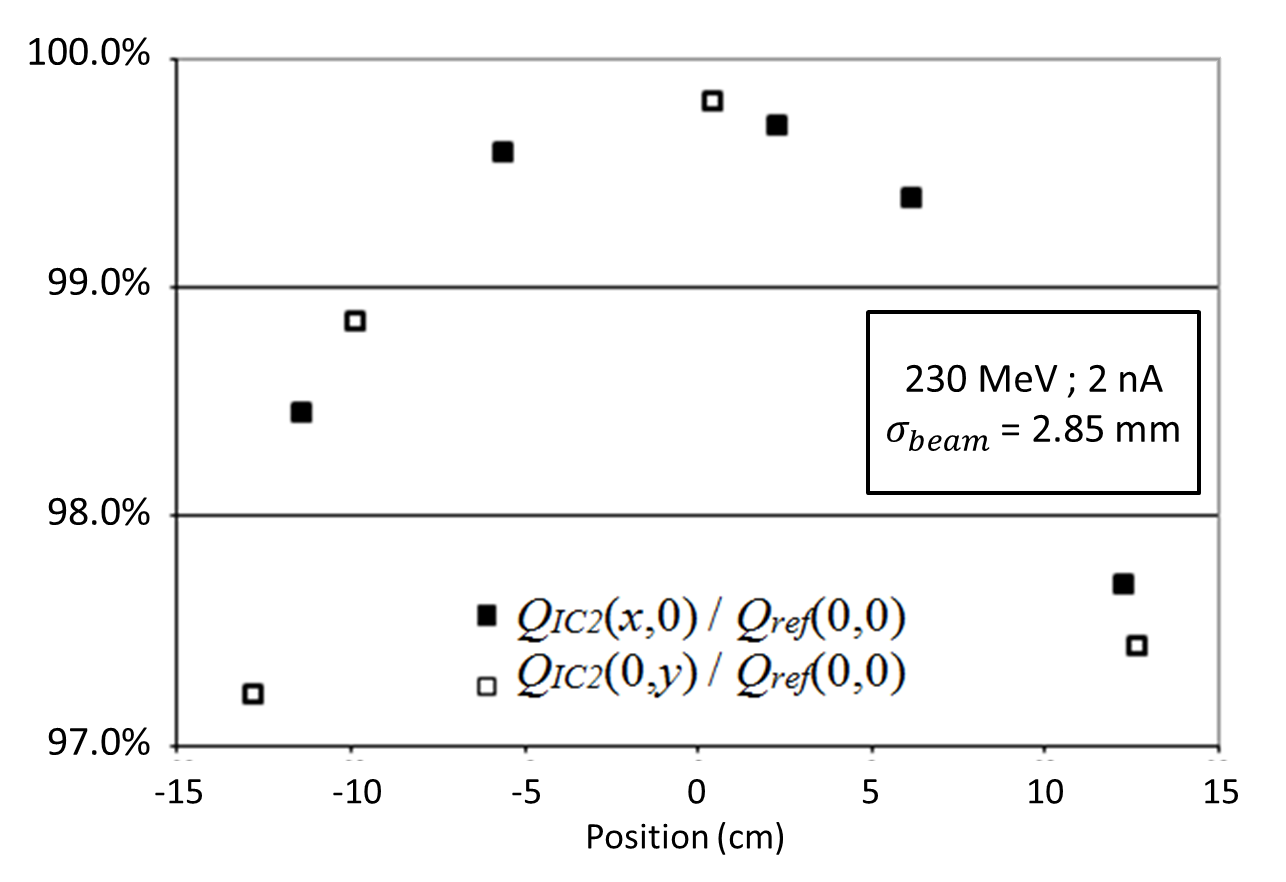}
\end{center}
\caption{ Uniformity test results.}
\label{Uniformity}
\end{figure}

\begin{figure}[!ht]
\begin{center}
\includegraphics[angle=0, width=\columnwidth]{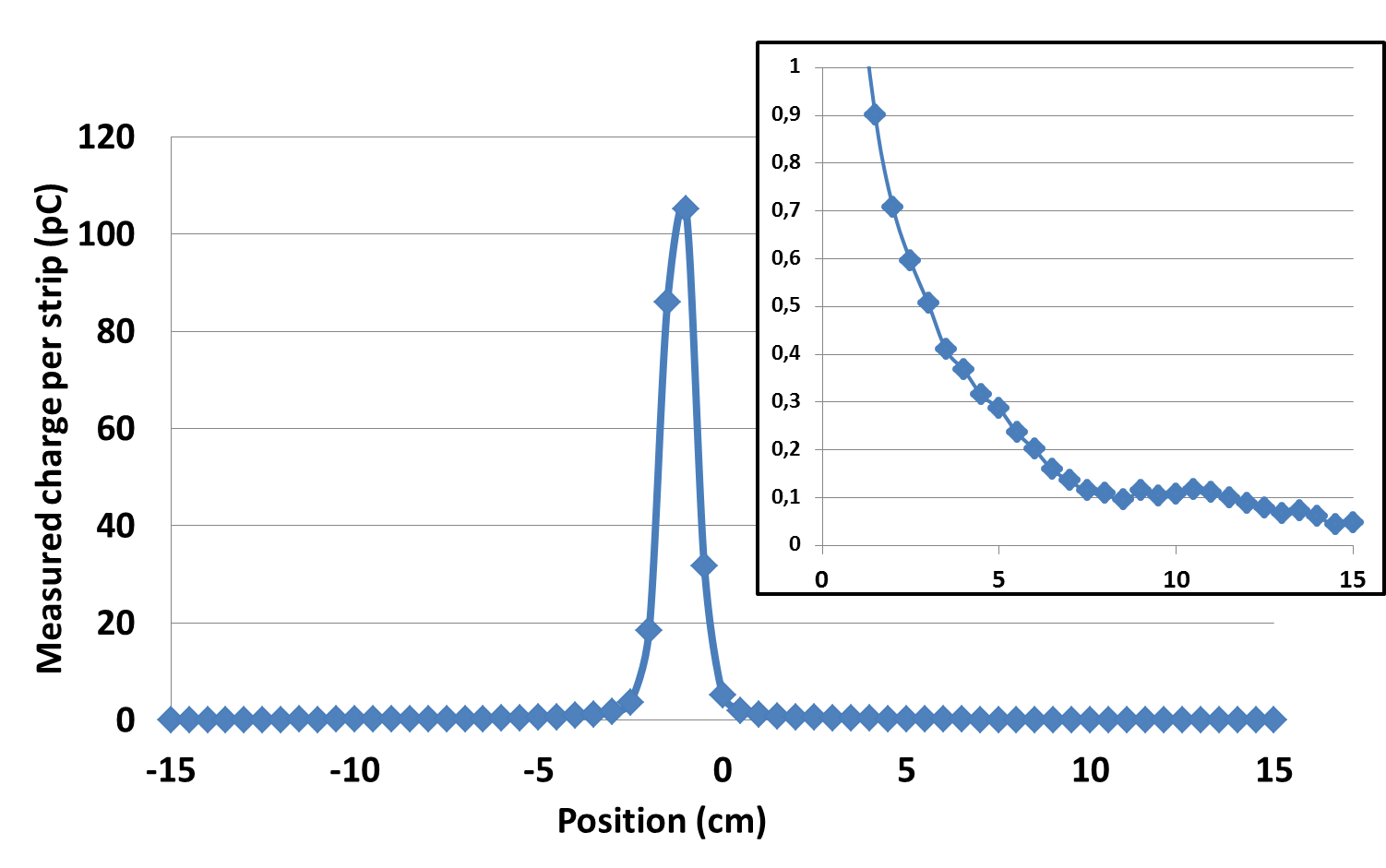}
\end{center}
\caption{Beam transverse profile on x axis.}
\label{BeamProfil}
\end{figure}

\begin{figure}[!ht]
\begin{center}
\includegraphics[angle=0, width=\columnwidth]{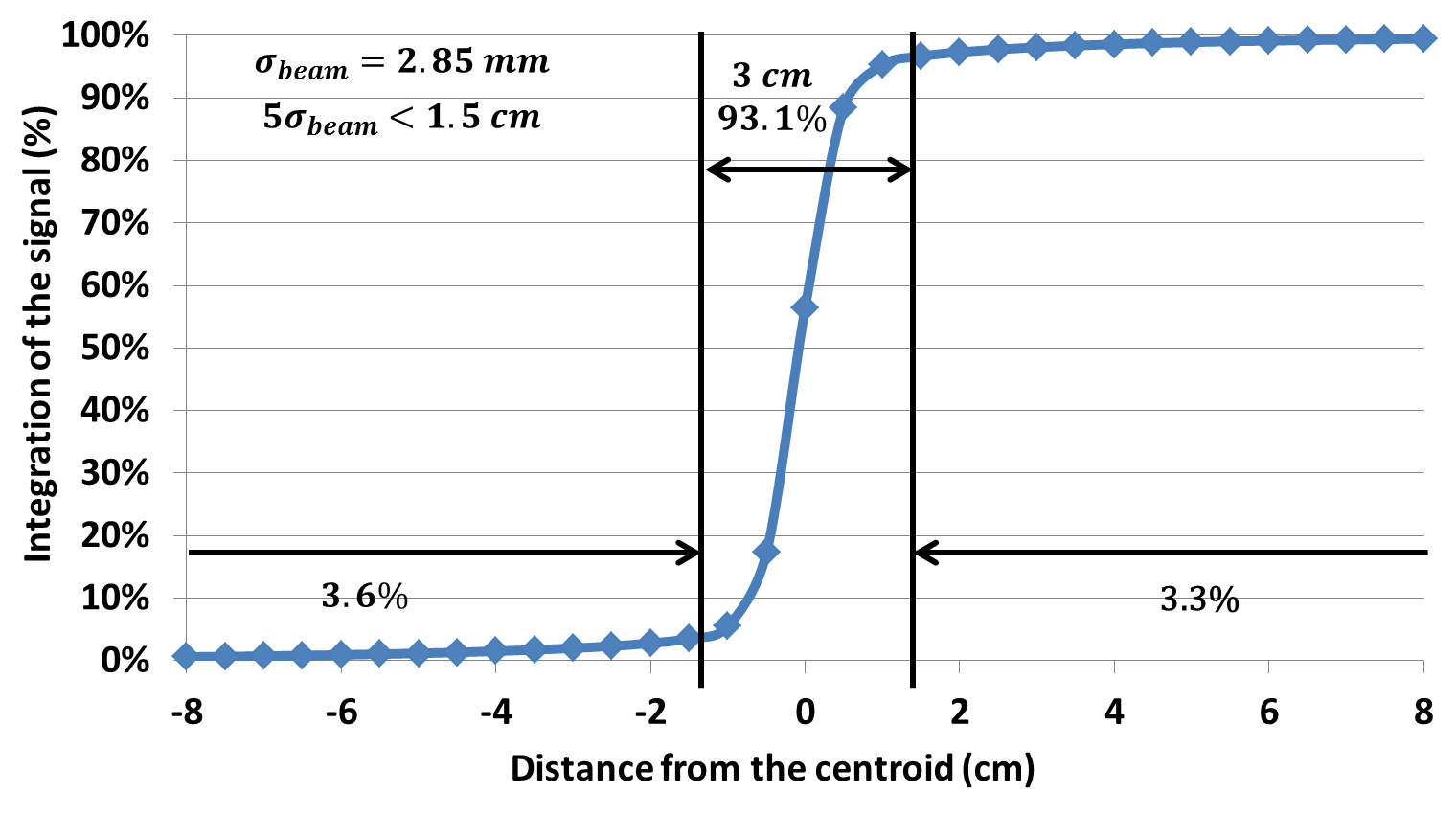}
\end{center}
\caption {Integrated dose measured by IC2/3 from $strip_0$ to $strip_i$ normalized by the measurement when the centroid is at the center of IC2/3.}
\label{BeamIntegration}
\end{figure}

Results obtained measuring the charge collected in different points on both $x$ and $y$ axes are presented in Fig.~\ref{Uniformity}. They show a 4\% variation in the charge collection between the center and the edges of the chamber. This can be explain by the measured dose profile which is not only Gaussian (see Fig.~\ref{BeamProfil}). In fact, the profile is the sum of a Gaussian shape and an exponential background. When the measurement position is the edge of the chamber half of the charges contained in those wings are lost and those does explain the 4\% of charge missing on the edge. Indeed, Fig.~\ref{BeamIntegration} shows the signal integrated over the 3~cm around the centroid (about five $\sigma_{beam}$ on each side) only contains 93.1\% of the charges.

To compensate those missing charges when the beam is close to the chamber edge, correction have be made. Those are linked to the amount of charge integrated by the detector as a function of the distance between the charge deposition centroid and the chamber center.
 
 \begin{figure}[!ht]
\begin{center}
\includegraphics[angle=0, width=\columnwidth]{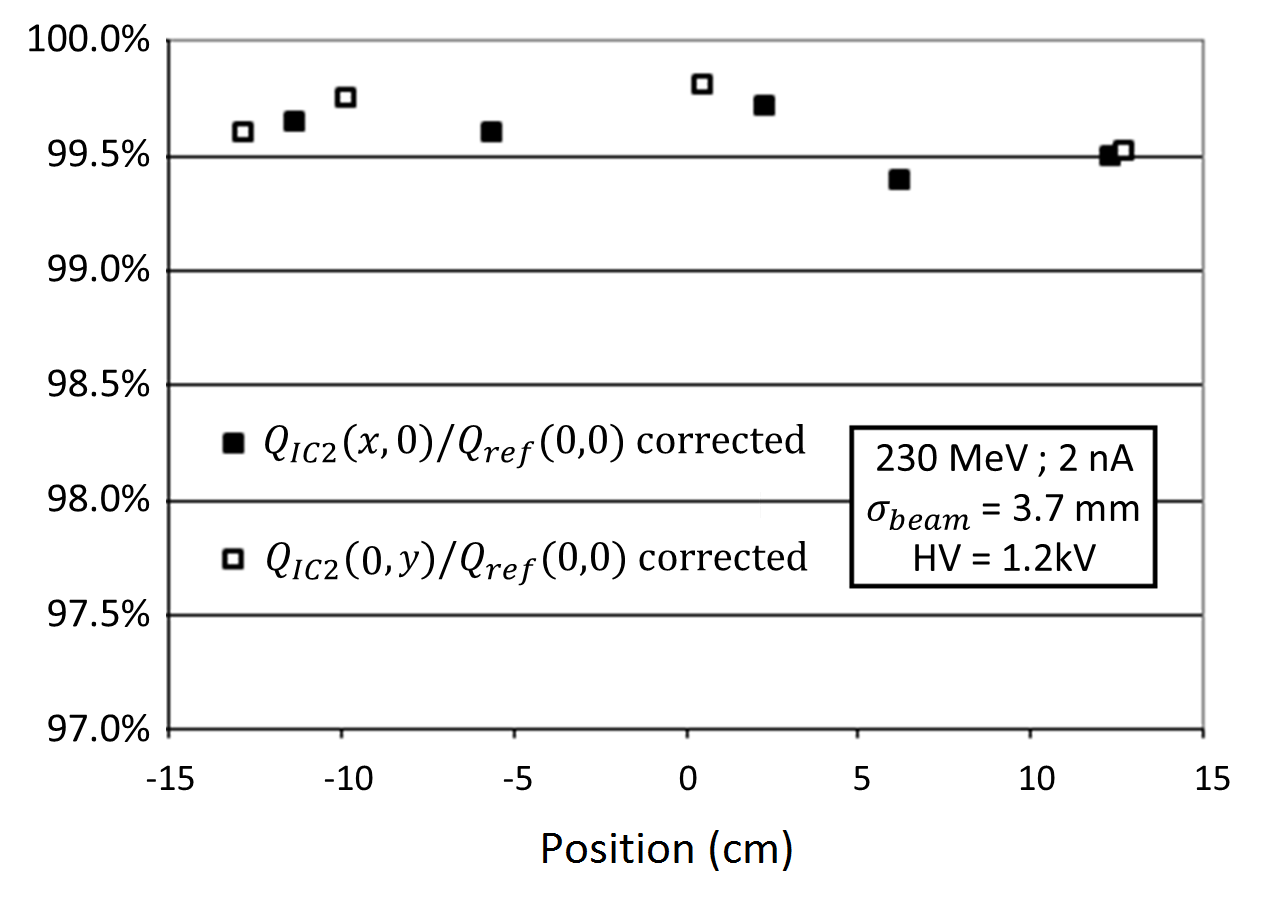}
\end{center}
\caption{Measured dose after correction for several centroid positions.}
\label{UniformityCorrected}
\end{figure}

Results of such calibration are presented on Fig.  \ref{UniformityCorrected}. This calibration allows to obtain a uniformity better than 1\%.

\subsection {Spatial Resolution}

\begin{figure}[!ht]
\begin{center}
\includegraphics[angle=0, width=\columnwidth]{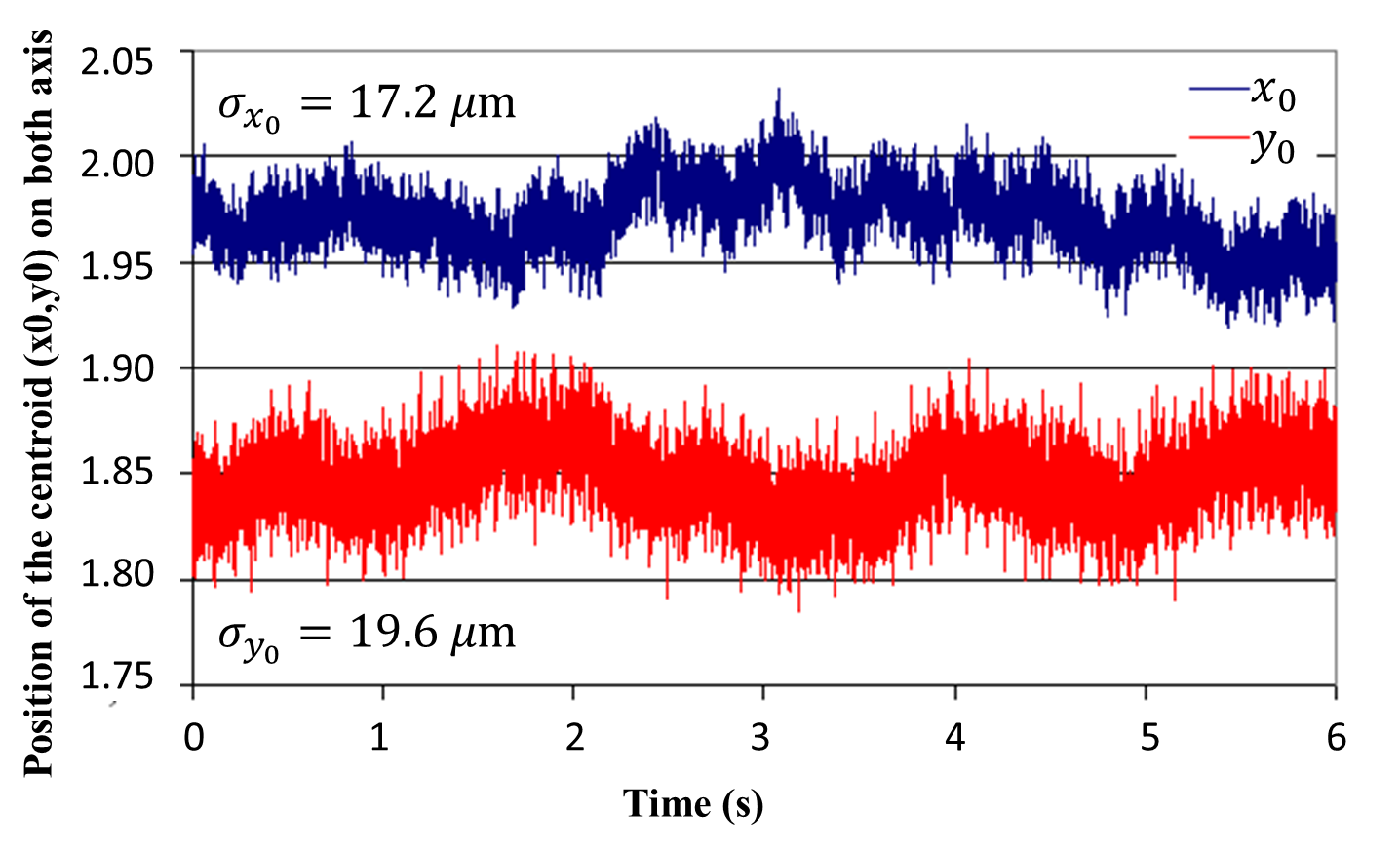}
\end{center}
\caption{Centroid measurements for a fixed beam as a function of time.}
\label{CentroidMeasurments}
\end{figure}

Fig.~\ref{CentroidMeasurments} shows the position of the centroid on the strip it belongs to as a function of time for a fixed beam so that the impact of the noise on the centroid estimate can be evaluated. The standard deviation of the measurements ($\sigma_{x_0}$ and $\sigma_{y_0}$) are around 20~$\mu$m with a peak-to-peak amplitude of 125~$\mu$m.

\begin{figure}[!ht]
\begin{center}
\includegraphics[angle=0, width=\columnwidth]{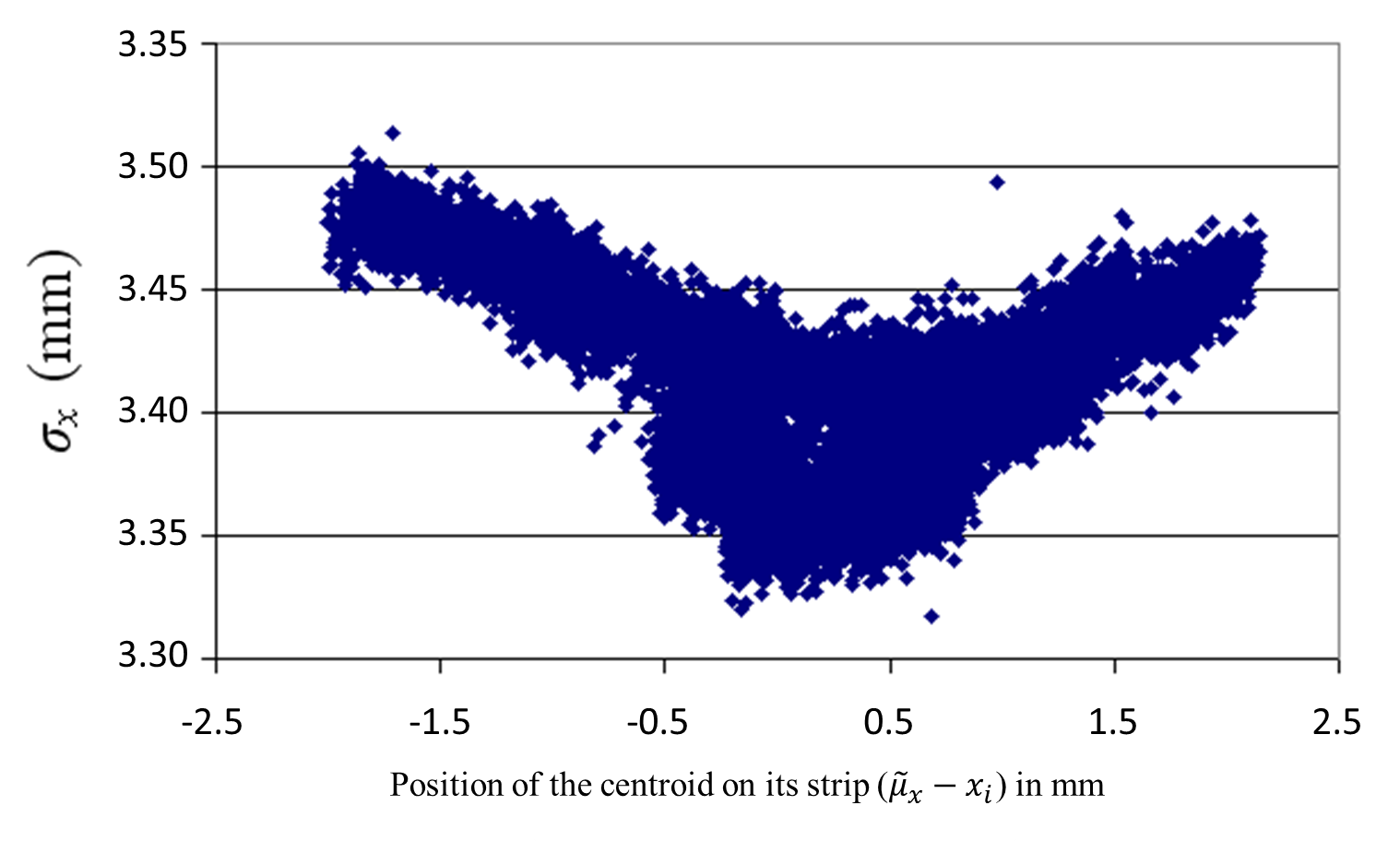}
\end{center}
\caption{ $\sigma_{beam}$ calculation as a function of the centroid relative position on its strip ($|\tilde{\mu}_x-x_i|$).}
\label{StandardDeviation}
\end{figure}

Fig.~\ref{StandardDeviation} shows the measured standard deviation of the $\sigma_{beam}$ value as a function of the position of the centroid on the strip it belongs for a scanning beam. The standard deviation of the $\sigma_{beam}$ value is 35~$\mu$m with a peak-to-peak value of 200~$\mu$m.

\section{Conclusion and perspectives}

The tests realized in the WPE institute in Essen showed that the IC2/3 chamber has the ability to measure a proton fluency with a 1\% precision in the 0.5-8~Gy/min range. In addition, we get a spatial resolution of the beam centroid of 20~$\mu$m and a time resolution of 500~$\mu$s. This chamber has been designed to have little impact on the beam due to its  187~$\mu$m water equivalent thickness and is to be used with a electrodes high voltage ($>$~1.2~kV) to ensure a collection efficiency better than 99.5\%. 
The chamber test results (Table ~\ref{table:Results}) meet the requirements for the Pencil Beam Scanning monitoring. Nowadays, five units of the IC2/3 chamber are installed in the WPE Essen protontherapy center and they are part of the Proteus 235 IBA commercial product.
The beam profile measurement shows a Gaussian shape with some wings on the edges. Those wings are not understood at the moment and even if they do not represent an issue for the dose measurement (if properly taken into account in the calibration) they need to be further investigated.

\begin{table}[!ht]
\begin{center}
\begin{tabularx}{\columnwidth}{ZZ}
\hline
\hline
Specification sheet&	Results\\
\hline
Equivalent water thickness&	187~$\mu$m\\
\hline
Time resolution	&500~$\mu$s (2~kHz sampling)\\
\hline
Repeatability&	$\sigma $/Q = 0.60\% (Table~\ref{table:Repeatability})\\
\hline
Collection efficiency	&$>$ 99.5\% in saturated regime ($>$~1.2~kV) (Fig.~\ref{SatCurve})\\
\hline
Linearity&	$\sigma $ = 0.24\% (Fig.~\ref{Linearity})\\
\hline
Response uniformity on the dose measurement layers (uniform film)	&1\% peak-to-peak between the center and the edges of the active area (Fig.~\ref{UniformityCorrected})\\
\hline
Spatial resolution	&20~$\mu$m (Fig.~\ref{CentroidMeasurments})\\
\hline
\hline
\end{tabularx}
\caption{Specifications and test results.}

\end{center}
\label{table:Results}
\end{table}

\begin{figure}[!ht]
\begin{center}
\includegraphics[angle=0, width=\columnwidth]{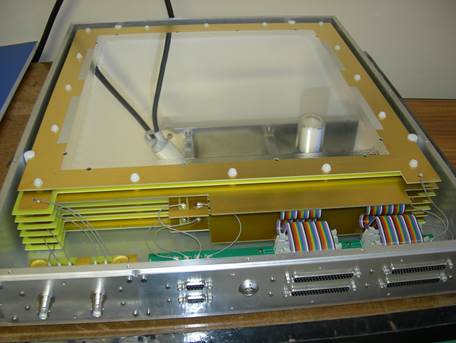}
\includegraphics[angle=0, width=\columnwidth]{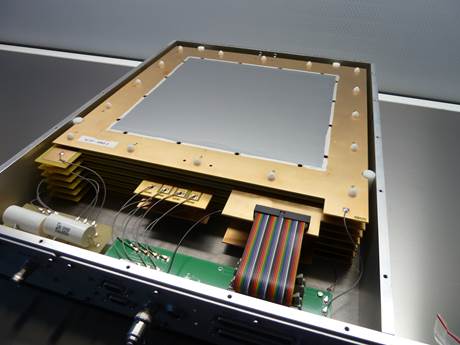}
\end{center}
\caption{Opened IC2/3 chamber (prototype in top and commercial in bottom.}
\label{ICopened}
\end{figure}


\begin{thebibliography}{99}
\bibitem{ICRU}ICRU Report 78 (2007).
\bibitem{Chu}W. T. Chu, B. A. Ludewigt and T. R. Renner, Review of Scientific Instruments 64 (1993).
\bibitem{Scardt}D. Schardt, T. Els\myacc{a}sser and D. Schulz-Ertner, Reviews of Modern Physics 82 (2010).
\bibitem{Lomax}A. J. Lomax and al., Medical Physics 31 (2004).
\bibitem{Marchand}B. Marchand, et al., Proceedings of EPAC 2000.
\bibitem{Giordanengo}S. Giordanengo, et al., Nuclear Instruments and Methods in Physics Research Section A (2013).
\bibitem{Boag}J. W. Boag, International Journal for Radiation Physics and Chemistry (1969).
\end{thebibliography}
\end{document}